# Disentangling kinetics from thermodynamics in heterogeneous colloidal systems

Hamed Almohammadi[1], Sandra Martinek[1], Ye Yuan[1], Peter Fischer[1], Raffaele Mezzenga[1,2*]

[1]Department of Health Sciences and Technology, ETH Zurich, Zurich, Switzerland
[2]Department of Materials, ETH Zurich, Zurich, Switzerland
*Correspondence to: raffaele.mezzenga@hest.ethz.ch

**Abstract**

Nucleation and growth (N&G) - the emergence of a new phase within an initially homogeneous one - is one of the most important physical phenomena by which gas-liquid (GLPS), liquid-liquid (LLPS) and solid-liquid (SLPS) phase separation takes place. Accordingly, thermodynamics sets the asymptotic boundaries towards which the system must evolve, while kinetics tries to cope with it by imposing the transport rates at which phase separation is realized. In all heterogeneous colloidal systems observed in nature, the composition, shape, structure and ultimately physical properties result from the trade-off between thermodynamics and kinetics. In this work we demonstrate, by carefully selecting colloidal systems and controlling phase separation in microfluidic devices, that it becomes possible to go beyond N&G, disentangling kinetics effects from thermodynamics in composition, structure and physical properties of the final system. Using amyloid fibril and cellulose nanocrystal filamentous colloids for which the binodal curve defining the two-phase region in the phase diagram is given by two separate vertical lines, we extrude a solution set at one thermodynamic branch inside the other branch, realizing nematic or cholesteric droplets where the composition is set by thermodynamics, while the structure and morphology are defined by dynamic flow parameters. We demonstrate that departing from the N&G paradigm unveils new physical phenomena, such as orders of magnitude shorter timescales, a wider phase diagram and internal cholesteric structures that are not observable via conventional LLPS. We also show that by co-dispersing plasmonic gold nanoparticles within colloidal liquid crystalline droplets, our approach enables on-demand fabrication of multicomponent heterogeneous liquid crystals, enhancing their potential, and introducing original fundamental and technological directions in multicomponent structured fluids.

The occurrence of N&G is ubiquitous in the universe and central to many scientific disciplines ranging from physics to material science, biology, and medicine. In SLPS, examples include phenomena as general as the freezing of water into ice[1-2], solidification of a molten metal[3], or formation of crystals in biomineralization[4]. In GLPS, N&G controls the formation of gas bubbles from supersaturated liquids[5], such as $CO_2$ bubbles sparking from a freshly opened bottle of champagne. N&G also plays a vital role in LLPS observed in polymeric fluids and colloidal dispersions; *in vivo*, it is furthermore associated with important biological processes such as the formation of intracellular membraneless organelles[6-8]. In technology, controlling N&G may allow reducing the timescale of formation of new phases and designing new materials with desired composition, structure, shape, size and ultimately physical properties[9-14]; not surprisingly N&G is indeed a process relevant to many fields including food[15], electronic[14,16], photonic[17] and pharmaceutical[18] industries. Yet, controlling N&G is largely limited by kinetic factors, including the stochastic nature of nucleation and the finite transport rates ruling the growth phase[19], so that the time parameter becomes pervasive in most heterogeneous systems, intimately connecting thermodynamics and kinetics aspects.

Heterogeneous systems based on filamentous biological colloids[20-25] are a class of matter that undergoes N&G via a distinct phase separation mechanism[26-27] and bears both fundamental and technological significance[28]. In this system, as first shown by Onsager[29], the phase separation has a purely entropic origin and stems from the interplay between orientational entropy and excluded volume packing entropy - a phenomenon known as liquid-liquid crystalline phase separation (LLCPS)[26-27] to distinguish it from the more common LLPS trade-off between enthalpy and entropy. This fundamental thermodynamic difference leads to a dramatic change in their respective phase diagrams: while the binodal line separating the 2-phase from the 1-phase region in LLPS maintains a finite slope[30], in LLCPS the binodal line is represented by two perfectly vertical lines separating the 2-phase region, where isotropic (I) and nematic (N) phases coexist, from the I and N single phase regions at low and high colloid composition, respectively.[29] These two vertical binodal branches occur at the Onsager volume fractions, which for monodisperse rods of length $L$ and diameter $D$, would locate at $\phi_I = 3.34(D/L)$ and $\phi_N = 4.49(D/L)$, respectively.[29] In practice, for polydisperse rod systems these two exact limits depend on rod length distribution[31]; for the two systems considered, however, an estimation of these two branches using average contour length of the rods has been shown to closely predict the experimental values[32-33].

Within the 2-phase concentration region (Fig. 1a), phase separation via N&G promotes formation of microdroplets of N phase with high concentration and orientational order within an I phase with no orientational order and low concentration[20-25]. These microdroplets, called tactoids, feature a fascinating rich phase behvaiour[25,32] and, unlike ordinary spherical colloids that crystallize to a roughly spherical shape, assume spindle-like, prolate or oblate shapes[20-25,32] due to the subtle interplay between the vanishingly small interfacial tension and liquid crystalline anisotropic elastic tensor[25,34] (Fig. 1b). Upon growth of the tactoids with time, with N&G time of the order of minutes to days when mechanical disturbance is avoided, their self-selected shape and structure change[20-25,32-34], with volume-composition curves evolving via trajectories of varying slopes depending on the system of interest, but always confined within the two Onsager vertical asymptotes (Fig. 1a). These volume-composition curves, which only recently have been theoretically estimated for rod-like objects undergoing LLCPS[26], directly reflect kinetics effects as they are a consequence of finite transport rates: only for infinitely fast transport, would they perfectly overlap with the two vertical composition Onsager lines. Thus, differently from LLPS, the infinite slope of the two Onsager vertical asymptotes sets a precise limit to the final compositions of the two phases and open to the possibility of manipulating and controlling morphology and structure by relying exclusively on the two equilibrium asymptotic compositions. For example, this could allow disentangling the internal periodicity of cholesteric tactoids, also known as cholesteric pitch, from the tactoid size, a breakthrough which has remained elusive to date[25-26].

In what follows, using β-lactoglobulin amyloid fibrils (BLG) and sulfated cellulose nanocrystals (SCNC) as two model anisotropic colloidal systems, we go beyond N&G by separating kinetics effects from thermodynamics with tangible consequences in composition, morphology, shape, structure and size distribution of the ensued tactoids. We bypass the kinetic pathway of N&G, thus replacing the classical N&G time of up to few days for the formation of tactoids with an *induction time* of the order of few minutes only. We also show that going beyond N&G produces tactoids with different aspect ratios compared to those obtained classically and that cholesteric droplets can be generated at different volumes, while maintaining a constant pitch, a structural hallmark departing significantly from tactoids obtained by N&G. We finally demonstrate the generality of the approach by creating highly controllable negative tactoids (I within N), and multicomponent liquid crystalline droplets based on biocolloids and plasmonic gold nanorods.

BLG and SCNC are carefully selected as two examples of filamentous colloids undergoing LLCPS via volume-composition curves of different shapes: we expect mild and progressive increase in the trajectory slope for BLG (e.g. such as the red line in Fig. 1a), because for this system the periodicity of cholesteric tactoids is known to depend significantly on time and volume[25], a strong enrichment of longer fibrils in the nematic phase is known to occur during N&G[32] and relaxation dynamic of the internal structure follows a slow kinetics[35]; on the contrary, we expect steep slopes of the volume-composition curve rapidly approaching the Onsager binodal limits for SCNC (e.g. such as the green line in Fig. 1a), since for this system fractionation of long and short fibrils in the nematic and isotropic phase, respectively, is small compared to BLG during N&G (see Ref. 32 for the length distributions of isotropic and nematic phases of SCNC and BLG, prepared following the same N&G protocol as here)[32], the cholesteric pitch is constant with tactoid volume and the relaxation kinetics is fast[35].

In our experiments, we use solutions of macroscopically phase separated and equilibrated biocolloids, corresponding to the two Onsager compositions[29] (Fig. 1c; see also Methods). We then extrude the nematic phase, set at one thermodynamic Onsager composition ($\phi_N$), inside the isotropic phase, set at the other Onsager composition ($\phi_I$), in a two-phase coflowing stream (Fig. 1d) obtaining heterogeneous colloidal systems of controlled droplet sizes, which we then compare with droplets of identical sizes, but generated by phase separation via N&G (see Methods for concertation values of BLG and SCNC). A well-established co-flow microfluidic chip consisting of two co-axially aligned capillary tubes is used to enable three-dimensional co-axial flow of the two phases[36-37]. The solution in $\phi_N$ is injected into the inner cylindrical capillary tube with 0.6 mm and 1 mm inner and outer diameters, respectively. One end of capillary tube is shaped into tapered orifice with 0.025 mm inner and 0.042 mm outer diameters, respectively. The surrounding medium in $\phi_I$ is injected in the interstices between the inner capillary tube and the outer capillary tube, which is square with 1.05 mm dimension. It is observed that upon injection of the solution using syringe pumps, the nematic jet is formed first; downstream the instability appears on the surface of the jet, which ultimately leads to breakup of the jet into droplets at $l_{jet}$. The jet diameter $d_{jet}$ is controlled easily by changing the flow rate of the inner $q_{in}$ and outer fluid $q_{out}$. However, care should be taken, since due to the vanishingly low surface tension of the tactoids[25,38] ~$10^{-7}$-$10^{-6}$ N m$^{-1}$, there is an upper limit for the flow rate, above which the tactoids are not able to keep their equilibrium shape and structure. Another upper limit for the flow speed is when the orientational order is induced by the flow field resulting in isotropic-to-nematic transition in the medium phase as pointed by de Gennes[39]. We perform our experiments below the mentioned limits where $q_{in}$ was maximum 5 µL h$^{-1}$ and $q_{out}$ maximum 50 µL h$^{-1}$.

BLG tactoids with various shapes, sizes and structures are formed by simply controlling the flow parameters (see Fig. 1e-g, Supporting Fig. 1, Supporting Videos 1-6). At low $q_{in}/q_{out}$ ratio of $7\times10^{-5}$, a nematic jet with $d_{jet} = 7.0$ µm is generated, which upon breakup results in formation of homogenous tactoids with equivalent diameter $D_{equiv.}$ (= $2V^{1/3}$ where $V = r^2R$ is the scaled volume of tactoids with $R$ and $r$ major and minor axes of tactoids, respectively) of 14.4 ± 1.6 µm (Fig. 1e, Supporting Videos 2). At $d_{jet}$ of 12.3 µm, achieved at $q_{in}/q_{out} = 23\times10^{-5}$, we observe the formation of tactoids with larger volume with $D_{equiv.}$ of 28.5 ± 3.6 µm and bipolar structure (Fig. 1f, Supporting Videos 3). At higher $q_{in}/q_{out} = 335\times10^{-5}$, a nematic jet with $d_{jet} = 48.3$ µm is generated which upon breakup leads to formation of droplets ($D_{equiv.} = 94.5 \pm 7.3$ µm) with cholesteric internal structure (Fig. 1g, Supporting Videos 6).

When performing experiments with SCNC (see Methods), cholesteric tactoids with $D_{equiv.} = 73.6 \pm 10.3$ µm are formed similarly to cholesteric BLG tactoids (Fig. 1h). Strikingly, we observe the formation of cholesteric structure in SCNC system already within the nematic jet where the cholesteric pitch $P$ of the jet measured to be 12.3 µm, slightly lower than the one observed for the SCNC tactoids formed beyond N&G with $P = 14.6$ µm. This is different from BLG system, implying that nematic jet of the SCNC relaxes to cholesteric structure before breakup, which can be related to the fast relaxation dynamics of SCNC system compared to BLG[35].

Our experiments reveal how the interplay between fluid flow and thermodynamics rules the dynamics of tactoids formation, leading to a rich behavior in the formation of the tactoids from the nematic liquid crystalline jet (Fig. 2a). Upon formation of the nematic jet, driven by Rayleigh-Plateau instability, the jet is axi-symmetrically perturbed downstream. The perturbation on the nematic jet grows by time and ultimately leads to a breakup of the jet into a chain of elongated droplets. This is followed by shape and structural relaxation of the droplets to their equilibrium state driven by free energy minimization where both liquid crystalline anisotropic and isotropic free energy terms contribute (see below). Another remarkable feature of our approach is the time scale of the tactoids formation, i.e. the *induction time*, which, as opposed to the N&G pathway taking minutes to days, is in the order of few minutes only (Supporting Fig. 2 and Fig. 3, Supporting Videos 1-6). This noteworthy reduction in the induction time is rooted in the bypass of the classical N&G kinetic pathway, allowing forming tactoids of final composition, structure and dimension independently from the limiting parameters of the N&G[40], such as the thermodynamic energy barrier of nucleation and the finite transport rates of the growth phase.

To establish a quantitative description of the size of the tactoids formed from a nematic jet with given $d_{jet}$, we approximate the volume of the tactoids to be the volume contained in one wavelength $\lambda$ of the instability appearing in the nematic jet, $V = \pi d_{jet}^2 \lambda/4$, where, for simplicity, we refer to the undisturbed stage of the jet still in its transient cylindrical shape of diameter $d_{jet}$. We consider the most unstable mode of the Rayleigh-Plateau instability[36-37,41-42] which is dependent on $\eta_N/\eta_I$ with $\eta_N$ and $\eta_I$, respectively, the viscosity of nematic and isotropic phases; with $\eta_N/\eta_I \approx 0.5$ [38], we obtain $\lambda = 4.6 d_{jet}$ [42]. Thus, we compute $D_{equiv.} = 1.9 d_{jet}$, predicting well and without fitting parameters the linear relation found in the experiments, as shown in Fig. 2b. Note that, although the viscosity of the nematic phase changes depending on the director field orientation, for simplicity we use the zero shear viscosity values for both isotropic and nematic phases in our analysis.

The shape relaxation behavior of the elongated droplets to the equilibrium state following breakup, shows a first order exponential decay following a universal curve of $(D(t)-D_{equil.})/(D_{initial}-D_{equil.}) = \exp(-t/\tau_s)$, with $D(t)$, $D_{initial}$ and $D_{equil.}$ the length of the tactoid at a given time $t$, at the initial time and at equilibrium state, respectively (Fig 2c). The time $t$ is parametrized by the shape characteristic relaxation time $\tau_s$ that is determined by fitting $(D(t)-D_{equil.})/(D_{initial}-D_{equil.}) = \exp(-t/\tau_s)$ to every single tactoid shape relaxation data set (see Supporting Fig. 4). These results show similar relaxation behavior as in our recent work on artificially elongated tactoids[43]. Furthermore, once formed, all tactoids are oriented in flow direction, except for cholesteric BLG tactoids with four or higher number of bands, which are orientated almost perpendicular to flow direction, a behavior whose underpinning physics is not entirely clear yet, calling for further investigation (see Fig. 1e-h, Supporting Fig. 1).

The 3D tactoidal phase diagram capturing the size $V$, shape (aspect ratio $\alpha = R/r$) and structure of the BLG tactoids formed using our approach is presented in Fig. 2d. We can generate tactoids with various volumes, ranging from 1.0 µm³ to 169,491.4 µm³, aspect ratios and internal structures. This expands significantly the scope of microfluidics in liquid crystalline systems that was limited, previously, to confining a nematic phase inside another fluid (e.g. oil), resulting only in spherical droplets.[44-45] We find the tactoids generated with our approach fairly monodisperse despite the low interfacial tension involved, with a coefficient of the variance of 11% for tactoids with $D_{equiv.} = 14.4$ µm, formed by a jet with $d_{jet} = 7.0$ µm (Fig. 2e).

Next, we take advantage of the possibilities offered by this approach to form the classes of the tactoids known as negative tactoids or atactoids[46-48]. These tactoids are understood to be microdroplets of the isotropic phase within the nematic phase and have been poorly studied[46]. To this end, we use the same microfluidic system as shown in Fig. 1d, but we extrude the isotropic phase set at the Onsager composition $\phi_I$, inside the nematic phase set at $\phi_N$ (see Fig. 1a). We note that the birefringence of the surrounding nematic medium makes impossible to distinguish the embedded isotropic tactoids under crossed polarizers, and therefore, the isotropic phase was dyed[49] with Thioflavin T (ThT) before extrusion and imaged by confocal fluorescence microscopy (see Fig. 3a-b, Methods). A jet of the isotropic phase is generated with different diameters by changing the flow rate (Fig. 3c-d). A clear boundary between isotropic and nematic phase is observed. A series of isotropic tactoids inside the nematic medium is formed (Fig. 3e), where as 3D reconstructed images in Fig. 3f and Supporting Video 7 illustrate, formation dynamics show similar steps as in the case of nematic tactoids within the isotropic phase. The phase diagram of the negative tactoids, showing the size and shape range, is given in Fig. 3g; note that the tactoids with small sizes are not identified which may be due to very weak emitted fluorescence signal, making it undetectable. Yet, our results show that, unlike direct tactoids, negative tactoids of different size hold nearly constant aspect ratio over four decades of volumes, which we understand to be a consequence of the isotropic nature of the dispersed phase (i.e. the negative tactoids). Additionally, the aspect ratio of 1.25 ± 0.08 is preserved up to volumes well beyond $10^7$ μm$^3$, sizes at which direct tactoids typically exhibit an aspect ratio of 1 (see Figure 4a for comparison).

Going beyond N&G allows revealing the emergence of physical phenomena not observed before. To highlight this, we carry out a comprehensive physical analysis on the tactoids formed beyond N&G and compare them to control experiments achieved via phase separation realized by classical N&G. The control experiments are done by triggering phase separation via N&G in two different ways: either diluting a system initially set at $\phi_N$ into the coexistence region, or by allowing direct N&G of tactoids from a composition falling within the 2-phase region, but previously destabilized by vortexing (Fig. 4a). While all three pathways show similar behavior of decreasing $\alpha$ with an increase in volume, when going beyond N&G several new fundamental differences appear.

First, BLG homogeneous tactoids of smaller volumes become detectable bypassing N&G (Fig. 4a). This is a direct consequence of the energy-activated nucleation process, which sets a minimum observable size for nuclei of the tactoids to be stable[26], whereas extrusion of the Onsager branches extends, virtually, to vanishingly small sizes (see Fig. 4a). Secondly, homogeneous tactoids with smaller $\alpha$ appear, which is also inaccessible to the other two N&G paths. The reason is that $\alpha$ is directly linked to the average length constituent particles[33] which is higher for the tactoids formed by N&G compared to those formed with our approach, since N&G selectively fractionates longer from shorter fibrils during phase separation (which is particularly true in the BLG case[32]). Note that no change is observed over time in the structure of the tactoids generated beyond N&G, indicating that the tactoids are essentially forming already in the final equilibrium configuration (Fig. 4b).

Thirdly, it becomes now even possible to highlight differences in the internal structure of tactoids achieved via or beyond N&G. To this end, we explore the relation of the cholesteric pitch $P$ with the volume of the tactoids in both the BLG and SCNC case. Previously, it was observed that $P$ decreases for BLG cholesteric tactoids as the volume grows during N&G[25-26]. This arises from the change in the concentration of the tactoids during the growth phase, and it is generally known that $P$ decreases with an increase in the concentration of the building blocks in the solution[50-52]. Remarkably, for BLG cholesteric tactoids formed beyond N&G, we observe a pitch value of 46.1 µm which stays constant over a broad range of tactoids volume size (Fig. 4c-d, Supporting Fig. 5). In contrast, the same BLG tactoids produced by N&G feature, as in previous reports[25], a steady decrease of the pitch $P$ with tactoid volume over the same range of sizes, until reaching a plateau value of 16.7 µm for macroscopically separated tactoids. The remarkable difference in pitch size observed in the two cases and the size-independence of $P$ when N&G is avoided, point at 46.1 µm as the true equilibrium pitch, which would then be only observable bypassing N&G; a plausible explanation is that the N&G process greatly enriches in longer fibrils the nematic phase, leading to a reduction of the pitch, which is known to decrease with the length of the colloids[33]. This shows the possibility to control the internal structure of tactoids beyond what is possible with N&G and allows us to experimentally confirm, for the first time, the theoretical rationale which has been advanced to explain the dependence between internal periodicity and volume of cholesteric tactoids during their growth[26].

Since the deviation between equilibrium and morphologies observed by N&G is understood to arise from the gentle progressive increase of the slope of the volume-composition curve towards the Onsager boundaries, we expect a completely different behavior when the trajectory is steeply and rapidly approaching the two asymptotic boundaries; in line with our expectation, and in stark contrast with BLG case, the SCNC cholesteric pitch remains constant, with no change in the internal structure of the tactoids over the same range of volume considered, and this independently of whether the tactoids are formed via or beyond N&G (Fig. 4e, Supporting Fig. 5): the proximity between the volume-composition curves and the Onsager asymptotes smear in the case of SCNC the composition and structural differences in the tactoids obtained by the two methods. Accordingly, the pitch of the SCNC tactoids formed by or beyond N&G is matching within the experimental error (Supporting Fig. 5) and found totally independent of time (Fig. 4e).

The possibility of controlling structure of heterogeneous liquid crystals beyond the state of the art, can also be extended to the generation of complex multicomponent colloidal fluids. As an example, we fabricate hybrid amyloid fibril-plasmonic gold nanorod tactoids by extruding a nematic phase (set at the binodal $\phi_N$) that is mixed with guest gold nanorods, in an isotropic medium (at binodal line $\phi_I$) (Fig. 4f-i). The nanorods follow the alignment of the host fibrils liquid crystal in the nematic jet and the ensuing tactoids, as evident in the fluorescence images with polarization sensitivity (Fig. 4i). The strong fluorescence signals in the images in Fig. 4g-i also show that the concentration of the nanorods remains high when the tactoids are formed[53]. This technique produces the hybrid tactoids within minutes, remarkably faster than the days needed by the existing methods[53-54], and allows full control over the concentration of the guest particles within the host amyloid tactoids, which is critically important for plasmonic applications, for enhancing fluorescence of amyloid fibrils, or for controlling the alignment and spatial distributions of gold nanorods within guest amyloid fibrils or cellulous nanocrystals[53-54].

In closing, we have shown that it is possible to bypass the kinetic pathway of N&G and disentangle kinetics from thermodynamics effects in heterogeneous colloidal systems. We have shown that our approach offers original possibilities in the formation of liquid crystalline droplets with significant reduction in induction time, expanded phase diagram, and control over their composition, morphology, shape, structure and size distribution. We have further demonstrated that it is possible to form on demand negative tactoids, that is, isotropic droplets in a nematic medium, and multicomponent liquid crystalline droplets with precision control of their size and composition. These findings deepen our fundamental understanding of the interplay between kinetics and thermodynamic in the formation of heterogeneous colloidal systems and may offer unexplored possibilities in all those technological applications relying on N&G as a sole mechanism to establish the final structure and morphology.

**Methods**

**Preparation of suspensions of amyloid fibrils.** The amyloid fibrils suspension was prepared from β-lactoglobulin purified from whey protein (see detailed protocol in Ref. 32). A suspension of 2 wt.% β-lactoglobulin in 300 mL Milli-Q water was prepared. The suspension was filtered through a 0.45 µm nylon syringe filter (Huberlab) and the pH of the suspension was adjusted to 2 by adding HCl. To form amyloid fibrils, the solution was then heated at 90 °C for 5 hours. This was followed by applying mechanical shear forces to shorten the length of the fibrils. To remove unreacted monomers and peptides from the suspension, dialysis was run for 5 days using a 100 kDa (MWCO) Spectra/Por dialysis membrane (Biotech CE Tubing) with daily bath replacement. The desired concentration for the suspension was achieved by up-concentrating with reverse osmosis method using a Spectra/Por 1 dialysis membrane (standard RC tubing) against a 10 wt.% polyethylene glycol solution (Sigma Aldrich) pH 2. After the complete macroscopic phase separation, the concentrations of the two Onsager branches were measured to be $\phi_I = 2.0 \pm 0.1$ wt% and $\phi_N = 2.5 \pm 0.2$ wt%.

**Preparation of suspension of cellulose nanocrystals.** As described in Ref. 32, 2.5 wt.% of freeze-dried cellulose nanocrystal (FPInnovations) was dissolved in milliQ water and sonicated for 120 seconds. To remove unwanted particles, the solution was centrifuged at 12,000 rcf for 20 min. For SCNC, we measured the concentrations of the two Onsager branches to be $\phi_I = 2.4 \pm 0.1$ wt% and $\phi_N = 2.9 \pm 0.2$ wt%.

**Microfluidic device.** The microfluidic device consisted of the two coaxially aligned capillary tubes: A square borosilicate glass capillary tube (VitroCom) with 1.05 mm inner dimension and a cylindrical borosilicate glass capillary tube (Hilgenberg) with inner and outer diameters of 0.60 mm and 1.00 mm, respectively. One end of the cylindrical capillary was shaped into a tapering orifice with inner and outer diameters of 0.025 mm and 0.042 mm at orifice tip, respectively. The round capillary was inserted into the square capillary. The capillaries were assembled on a glass slide (Thermo Scientific) with the help of Bondic liquid plastic welder.

One end of the inner and the outer capillary tubes were connected to a blunt needle with 0.34 mm inner diameter and 0.64 mm outer diameter. A flexible tube with 0.8 mm inner diameter is used to connect the syringes (Hamilton), filled with the solutions, to the blunt needles. Syringe pumps (PhD 2000, Harvard Apparatus) are used to inject the solutions into the microcapillary device.

**Experimental details.** To ensure that there were no contaminants in the system, all capillaries, tips and syringes were cleaned with ethanol and to avoid tip blockage in the capillary tubes, the solutions were centrifuged before each experiment. In our experiments, great care was taken to prevent backflow, which may lead to mixture of two phases. For instance, when starting the experiments, the syringe pumps were run in a way to move the front edge of both jet and the medium phase at the same time; later the flow rates were adjusted to the desired rate. All experiments were performed at room temperature.

**Sample characterization.** Four different microscopy systems were used to characterize different types of samples involved. First, cross-polarized optical microscopy (Zeiss Axio Imager Z2) with an attached camera (AxioCam MRc) with 5x (Achrostigmat), 10x (Plan Neofluar), 20x (Epiplan-Neofluar) and 50x (Epiplan-Neofluar) objectives was used to record the images of the tactoids. The images were acquired under time-series mode at frame rates between 300 to 600 frames per minute depending on the speed of the flow inside the microfluidic system. Note that the microfluidic device was placed with 45 ° with respect to one of the polarizers of the microscope, allowing us to detect the jet and tactoids clearly. Second, a microscope (Zeiss Axio Imager M1m) equipped with LC (liquid crystal)-PolScope and objectives of 5x (Achrostigmat), 10x (Plan Neofluar), and 20x (Epiplan-Neofluar) was used to capture the PolScope images of the tactoids. Third, confocal microscopy (Zeiss LSM 780 Axio Imager 2) with 10x objective (EC Plan-Neofluar) was used to record the results reported in Figure 3. All analyses were performed using the ImageJ and/or Zen software. Forth, in the experiments of hybrid amyloid fibrils-plasmonic gold nanorods tactoids, a confocal microscope (Leica TCS SP8) with a 10× objective (HC PL Fluotar Ph1) is used to acquire the polarization-dependent fluorescence images. The excitation source used in our experiments was a continuous laser at 633 nm with the power of 4 mW,

**Preparation of amyloid fibrils dyed with ThT.** In the case of the negative tactoids experiments, the isotropic phase was stained with the fluorophore Thioflavin T (ThT) with mix ratio of 100:1 by volume. ThT binds specifically to the β-sheets of the amyloid fibers and shows enhanced fluorescence. In our microscopy measurements, the excitation was set at 458 nm wavelength and the detection wavelength between 463 to 553 nm.

**Gold nanorods experiments.** Gold nanorods are prepared following the protocol described in Ref. 50. In our experiments, we mixed 40 μL of the highly concentrated GNR dispersion with 200 μL of the nematic phase of amyloid fibrils solution. Then, the mixture was filled in the co-flow microfluidic chips and imaged with fluorescence microscopy as described above.

**Acknowledgments**

We thank T. Schwarz and S. S. Lee from the Scientific Center for Optical and Electron Microscopy of ETH Zurich (ScopeM) for the help with the polarized optical microscopy and the laser scanning confocal microscopy and for the technical supports with microfluidic experiments. We also thank Q. Sun, I. Kutzli, J. Zhou, and J. Vogt for technical support and discussions. Support from the Swiss National Science Foundation - Sinergia Scheme Grant No. CRSII5_189917 - is gratefully acknowledged.

**Author Contributions**

H.A. and R.M. conceived the idea and initiated the project, designed the experiments, and performed theoretical interpretation of the results and the theoretical analysis underpinning the project. H.A. and S.M. built the experimental apparatus, performed experiments, and analyzed the data. H.A. and Y.Y. performed gold nanorods experiments and analyzed the data. P.F. contributed to the experiments. R.M. supervised the research. H.A., S.M. and R.M. wrote the manuscript with the input from all authors.

**References**

1. Bartels-Rausch, T. et al. Ice structures, patterns, and processes: a view across the icefields. *Rev. Mod. Phys.* **84**, 885–944 (2012).

2. Matsumoto, M., Saito, S. & Ohmine, I. Molecular dynamics simulation of the ice nucleation and growth process leading to water freezing. *Nature* **416**, 409–413 (2002)

3. Boettinger, W. J. et al. Solidification microstructures: recent developments, future directions. *Acta Mater.* **48**, 43–70 (2000).

4. Van Driessche, A. E., Kellermeier, M., Benning, L. G. & Gebauer, D. (Eds.). *New perspectives on mineral nucleation and growth: from solution precursors to solid materials*. (Springer, 2016).

5. Blander, M. & Katz, J. Bubble nucleation in liquids. *AIChE J.* **21**, 833–848 (1975).

6. Berry, J., Brangwynne, C. P. & Haataja, M. Physical principles of intracellular organization via active and passive phase transitions. *Rep. Prog. Phys.* **81**, 046601 (2018).

7. Alberti, S., Gladfelter, A. & Mittag, T. Considerations and challenges in studying liquid-liquid phase separation and biomolecular condensates. *Cell* **176**, 419–434 (2019).

8. Hyman, A. A., Weber, C. A. & Jülicher, F. Liquid–liquid phase separation in biology. *Annu. Rev. Cell Dev. Biol.* **30**, 39–58 (2014).

9. Dupuy, J., Jal, J. F., Ferradou, C., Chieux, P., Wright, A. F., Calemczuk, R. & Angell, C. A. Controlled nucleation and quasi-ordered growth of ice crystals from low temperature electrolyte solutions. *Nature* **296**, 138-140 (1982).

10. Aizenberg, J., Black, A. J. & Whitesides, G. M. Control of crystal nucleation by patterned self-assembled monolayers. *Nature* **398**, 495-498 (1999).

11. Liz-Marzán, L. M. & Grzelczak, M. Growing anisotropic crystals at the nanoscale. *Science* **356**, 1120–1121 (2017).

12. Yin, Y. & Alivisatos, A. P. Colloidal nanocrystal synthesis and the organic–inorganic interface. *Nature* **437**, 664–670 (2005).

13. Diao, Y., Harada, T., Myerson, A. S., Hatton, T. A. & Trout, B. L. The role of nanopore shape in surface-induced crystallization. *Nature Mater.* **10**, 867–871 (2011).

14. El-Sayed, M. A. Small is different: shape-, size-, and composition-dependent properties of some colloidal semiconductor nanocrystals. *Acc. Chem. Res.* **37**, 326–333 (2004).

15. Dickinson, E. in *Food Colloids. Proteins, Lipids and Polysaccharides* (eds Dickinson, E. & Bergenstahl, B.) 107–126 (Royal Society of Chemistry, Cambridge, 1997).

16. Talapin, D. V., Lee, J-S., Kovalenko, M. V. & Shevchenko, E. V. Prospects of colloidal nanocrystals for electronic and optoelectronic applications. *Chem. Rev.* **110**, 389–458 (2009).

17. Shirasaki, Y., Supran, G. J., Bawendi, M. G. & Bulović, V. Emergence of colloidal quantum-dot light-emitting technologies. *Nature Photon.* **7**, 13–23 (2013).

18. Cox, J. R., Ferris, L. A. & Thalladi, V. R. Selective growth of a stable drug polymorph by suppressing the nucleation of corresponding metastable polymorphs. *Angew. Chem. Int. Ed.* **46**, 4333–4336 (2007).

19. Thanh, N. T., Maclean, N. & Mahiddine, S. Mechanisms of nucleation and growth of nanoparticles in solution. *Chem. Rev.* **114**, 7610–7630 (2014).

20. Zocher, H. Über freiwillige Strukturbildung in Solen. (Eine neue Art anisotrop flüssiger Medien.). *Zeitschrift für anorganische und allgemeine Chemie* **147**, 91-110 (1925).

21. Bawden, F. C., Pirie, N. W., Bernal, J. D. & Fankuchen, I. Liquid crystalline substances from virusinfected plants. *Nature* **138**, 1051–1052 (1936).

22. Dogic, Z. Surface freezing and a two-step pathway of the isotropic-smectic phase transition in colloidal rods. *Phys. Rev. Lett.* **91**, 165701 (2003).

23. Puech, N., Grelet, E., Poulin, P., Blanc, C. & Van Der Schoot, P. Nematic droplets in aqueous dispersions of carbon nanotubes. *Phys. Rev. E* **82**, 020702 (2010).

24. Revol, J.-F., Bradford, H., Giasson, J., Marchessault, R. H. & Gray, D. G. Helicoidal self-ordering of cellulose microfibrils in aqueous suspension. *Int. J. Biol. Macromol.* **14**, 170–172 (1992)

25. Nyström, G., Arcari, M. & Mezzenga, R. Confinement-induced liquid crystalline transitions in amyloid fibril cholesteric tactoids. *Nat. Nanotech.* **13**, 330 (2018).

26. Azzari, P., Bagnani, M. & Mezzenga, R. Liquid–liquid crystalline phase separation in biological filamentous colloids: nucleation, growth and order–order transitions of cholesteric tactoids. *Soft Matter* **17**, 6627-6636 (2021).

27. Fraccia, T. P. & Zanchetta, G. Liquid–liquid crystalline phase separation in biomolecular solutions. *Curr. Opin. Colloid Interface Sci.*, 101500 (2021).

28. Smalyukh, I. I. Liquid crystal colloids. *Annu. Rev. Condens. Matter Phys.* **9**, 207–226 (2018).

29. Onsager, L. The effect of shape on the interaction of colloidal particles. *Ann. NY Acad. Sci.* **51**, 627–659 (1949).

30. de Gennes, P.-G. *Scaling Concepts in Polymer Physics* (Cornell Univ. Press, 1979).

31. Honorato-Rios, C. & Lagerwall, J. P. F. Interrogating helical nanorod self-assembly with fractionated cellulose nanocrystal suspensions. *Commun. Mater.* **1**, 1–11 (2020).

32. Bagnani, M., Azzari, P., De Michele, C., Arcari, M. & Mezzenga, R. Elastic constants of biological filamentous colloids: estimation and implications on nematic and cholesteric tactoid morphologies. *Soft Matter* **17**, 2158-2169 (2021).

33. Bagnani, M., Nyström, G., De Michele, C. & Mezzenga, R. Amyloid fibrils length controls shape and structure of nematic and cholesteric tactoids. *ACS Nano* **13**, 591–600 (2019).

34. Prinsen, P. & van der Schoot, P. Shape and director-field transformation of tactoids. *Phys. Rev. E* **68**, 21701 (2003).

35. Khadem, S. A., Bagnani, M., Mezzenga, R. & Rey, A. Relaxation dynamics in bio-colloidal cholesteric liquid crystals confined to cylindrical geometry. *Nat. Commun.* **11**, 4616 (2020).

36. Cramer, C., Fischer, P. & Windhab, E. J. Drop formation in a co-flowing ambient fluid. *Chem. Eng. Sci.* **59**, 3045–3058 (2004).

37. Utada, A. S., Fernandez-Nieves, A., Stone, H. A. & Weitz, D. A. Dripping to jetting transitions in coflowing liquid streams. *Phys. Rev. Lett.* **99**, 094502 (2007).

38. Almohammadi, H., Bagnani, M. & Mezzenga, R. Flow-induced order–order transitions in amyloid fibril liquid crystalline tactoids. *Nat. Commun.* **11**, 5416 (2020).

39. de Gennes, P. G. & Prost, J. *The Physics of Liquid Crystals* (Clarendon Press, 1993).

40. Cuetos, A. & Dijkstra, M. Kinetic pathways for the isotropic-nematic phase transition in a system of colloidal hard rods: a simulation study. *Phys. Rev. Lett.* **98**, 095701 (2007).

41. Tomotika, S. On the instability of a cylindrical thread of a viscous liquid surrounded by another viscous fluid. *Proc. R. Soc. A* **150**, 322 (1935).

42. Cheong, A. G. & Rey, A. D. Temperature effects on capillary instabilities in a thin nematic liquid crystalline fiber embedded in a viscous matrix. *Eur. Phys. J. E* **9**, 171-193 (2002).

43. Almohammadi, H., Khadem, S. A., Bagnani, M., Rey, A. & Mezzenga, R. Shape and structural relaxation of colloidal tactoids. *Submitted*.

44. Parker, R. M. et al. Hierarchical self-assembly of cellulose nanocrystals in a confined geometry. *ACS Nano* **10**, 8443–8449 (2016).

45. Li, Y. et al. Colloidal cholesteric liquid crystal in spherical confinement. *Nat. Commun.* **7**, 12520 (2016).

46. Wang, P. X. & MacLachlan, M. J. Liquid crystalline tactoids: ordered structure, defective coalescence and evolution in confined geometries. *Philos. Trans. A Math Phys. Eng. Sci.* **376**, 20170042 (2018).

47. Watson, J. H. L., Heller, W. & Wojtowicz, W. Comparative electron and light microscopic investigations of tactoid structures in V2O5-sols. *Science* **109**, 274–278 (1949).

48. Oakes, P. W., Viamontes, J. & Tang, J. X. Growth of tactoidal droplets during the first-order isotropic to nematic phase transition of F-actin. *Phys. Rev. E* **75**, 061902 (2007).

49. Vassar, P. F. & Culling, C. S. Fluorescent stains, with special reference to amyloid and connective tissues. *Arch. Pathol.* **68**, 487–498 (1959).

50. Grelet, E. & Fraden, S. What is the origin of chirality in the cholesteric phase of virus suspensions? *Phys. Rev. Lett.* **90**, 198302 (2003).

51. Siavashpouri, M. et al. Molecular engineering of chiral colloidal liquid crystals using DNA origami. *Nat. Mater.* **16**, 849–856 (2017).

52. Dogic, Z. & Fraden, S. Cholesteric phase in virus suspensions. *Langmuir* **16**, 7820–7824 (2000).

53. Yuan, Y., Almohammadi, H., Probst, J. & Mezzenga, R. Plasmonic amyloid tactoids. *Adv. Mater.* **33**, 2106155 (2021).

54. Liu, Q., Campbell, M. G., Evans, J. S. & Smalyukh, I. I. Orientationally ordered colloidal co-dispersions of gold nanorods and cellulose nanocrystals. *Adv. Mater.* **26**, 7178-7184 (2014).

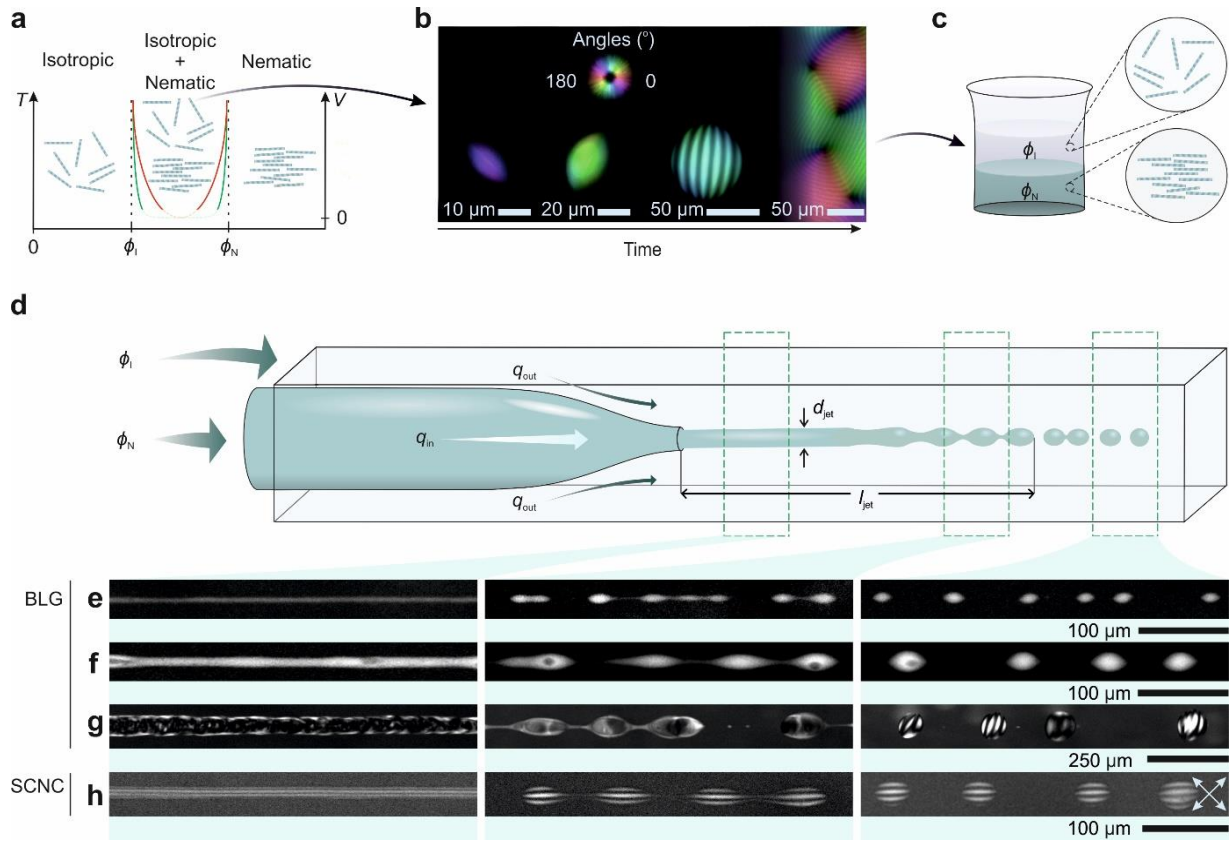

**Fig. 1 | Disentangling kinetics effects from thermodynamics in anisotropic bio-colloidal systems**.
**a**, Phase diagram overlaid with volume-composition diagram illustrating the two Onsager branches, $\phi_I$ and $\phi_N$, for anisotropic colloidal systems, along with volume-composition kinetic trajectories of tactoids growing by N&G. The red and green curves sketch, respectively, the volume-composition curves of the growth process of β-lactoglobulin amyloid fibrils (BLG) and sulfated cellulose nanocrystals (SCNC) systems with an initial concentration within $\phi_I$ and $\phi_N$. The vertical axes denote the temperature $T$ for the phase diagram and the volume $V$ for the emerging nematic phase (tactoids). The volume-composition curves are sketches only: precise trajectories have been theoretically derived elsewhere[26]. **b**, LC (liquid crystal)-PolScope images showing the formation of homogenous (first column), bipolar (second column), cholesteric tactoids (third column) and bulk phase (fourth column), through nucleation and growth (N&G), in a solution where the concentration is set within the coexistence region, isotropic + nematic (I+N). The colormaps (first row) show the orientation of the director field in the plane. **c**, Schematic of a completely phase separated solution initially set at a concentration within the coexistence region, where the nematic phase at concentration $\phi_N$ and the isotropic phase at concentration $\phi_I$ appear at the bottom and on top, respectively. **d**, Schematic of the co-flow microcapillary device used to extrude a solution set at one thermodynamic Onsager branch $\phi_N$ inside the other Onsager branch $\phi_I$. The tactoids are formed following jet breakup at $l_{jet}$. **e-h**, Cross-polarized images of the process of the formation of homogenous (**e**), bipolar (**f**), and cholesteric (**g**) tactoids with BLG and cholesteric tactoids of SCNC (**h**). To form tactoids with various shapes, volumes and structures, nematic jet with various diameters $d_{jet}$ (first column) are formed by changing the flow rate of the inner $q_{in}$ and outer fluid $q_{out}$, which is followed by breakup of the jet (second column) and formation of the tactoids (third column). The crossed arrows denote the crossed polarizers.

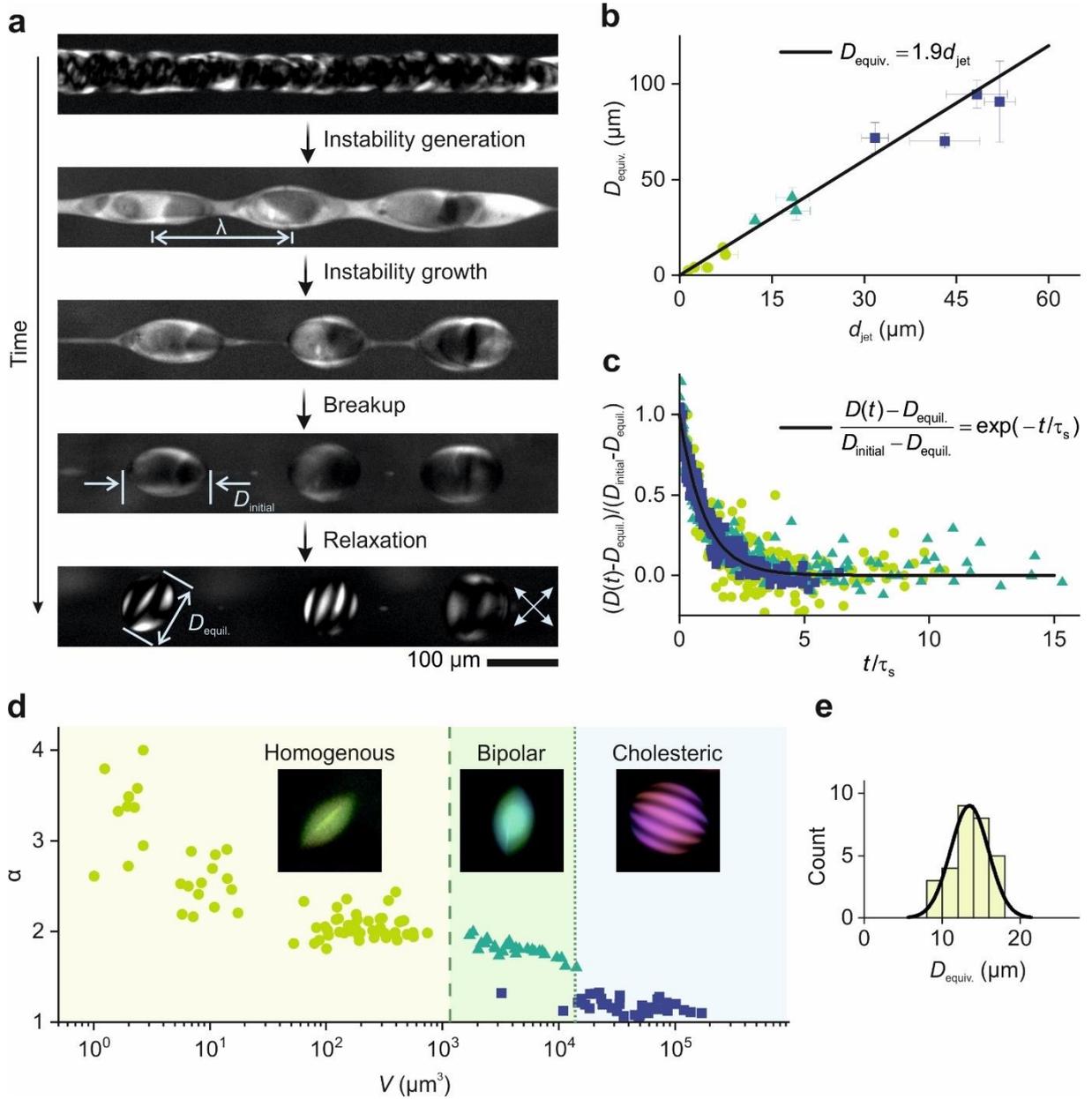

**Fig. 2 | The interplay between fluid flow and thermodynamics rules the dynamics of tactoids formation.** The filled circle, triangle, and square symbols in the plots represent data points from homogenous, bipolar, and cholesteric tactoids, respectively. **a**, Rich dynamics of formation of the tactoids captured under the crossed-polarizer: formation of nematic jet, development of Rayleigh-Plateau instability with wavelength λ, breakup of jet, and shape and structural relaxation of tactoids with initial length of $D_{initial}$ and equilibrium length of $D_{equil.}$. The crossed arrows denote the crossed polarizers. **b**, Experimental data of equivalent diameter $D_{equiv.}$ (=$2V^{1/3}$ where $V = r^2R$ is the scaled volume of tactoids with $R$ and $r$ major and minor axes of tactoids, respectively) as a function of the jet diameter $d_{jet}$. The prediction (solid line) is in good agreement, without fitting parameters, with experiments showing linear relationship between $d_{jet}$ and $D_{equiv.}$ for homogenous, bipolar and cholesteric tactoids. **c**, Relaxation behavior shown by evolution in $(D(t)-D_{equil.})/(D_{initial}-D_{equil.})$, with $D(t)$ the length of the tactoids at a given time, as a function of time $t$ parametrized by the shape characteristic relaxation time $\tau_s$. The results from the tactoids with various internal structures and volumes follow a universal curve: $(D(t)-D_{equil.})/(D_{initial}-D_{equil.}) = \exp(-t/\tau_s)$. **d**, The 3D nematic-cholesteric phase diagram showing the size $V$, shape (aspect ratio $\alpha = R/r$) and structure of BLG tactoids generated beyond N&G. **e**, Monodispersity of the formed homogeneous tactoids with $D_{equiv.} = 14.4$ μm showing a 11% coefficient of the variance.

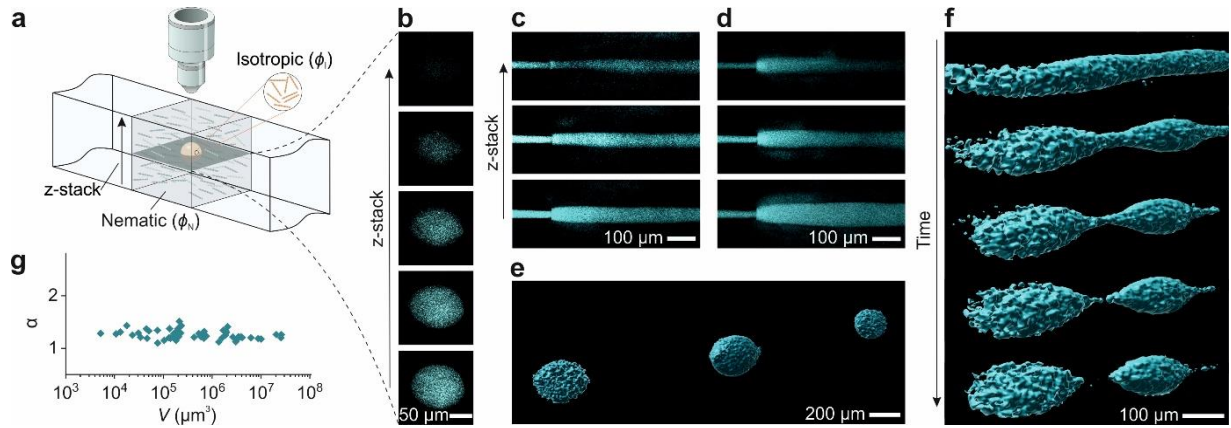

**Fig. 3 | Negative tactoids formation beyond N&G. a**, Schematic of an isotropic droplet, that is set at one thermodynamic Onsager branch $\phi_I$, within a nematic phase set at the other Onsager branch $\phi_N$, in the microfluidic system, where the fibrils in the isotropic phase have been fluorescently dyed with Thioflavin T (ThT). **b**, Confocal images of a single isotropic droplet in different z-stacks, showing the visibility of the tactoid at different heights. **c-d**, Confocal images of isotropic jet with different diameters close to the tip of the inner capillary tube at different z-stacks. **e**, 3D reconstructed image of three negative tactoids generated beyond N&G. **f**, 3D reconstructed images showing the dynamics of the negative tactoids formation: formation of the negative jet, appearance of the instability, breakup of the jet, relaxation of the negative tactoids. **g**, Phase diagram of the negative tactoids formed beyond N&G showing aspect ratio α of tactoids versus their volumes.

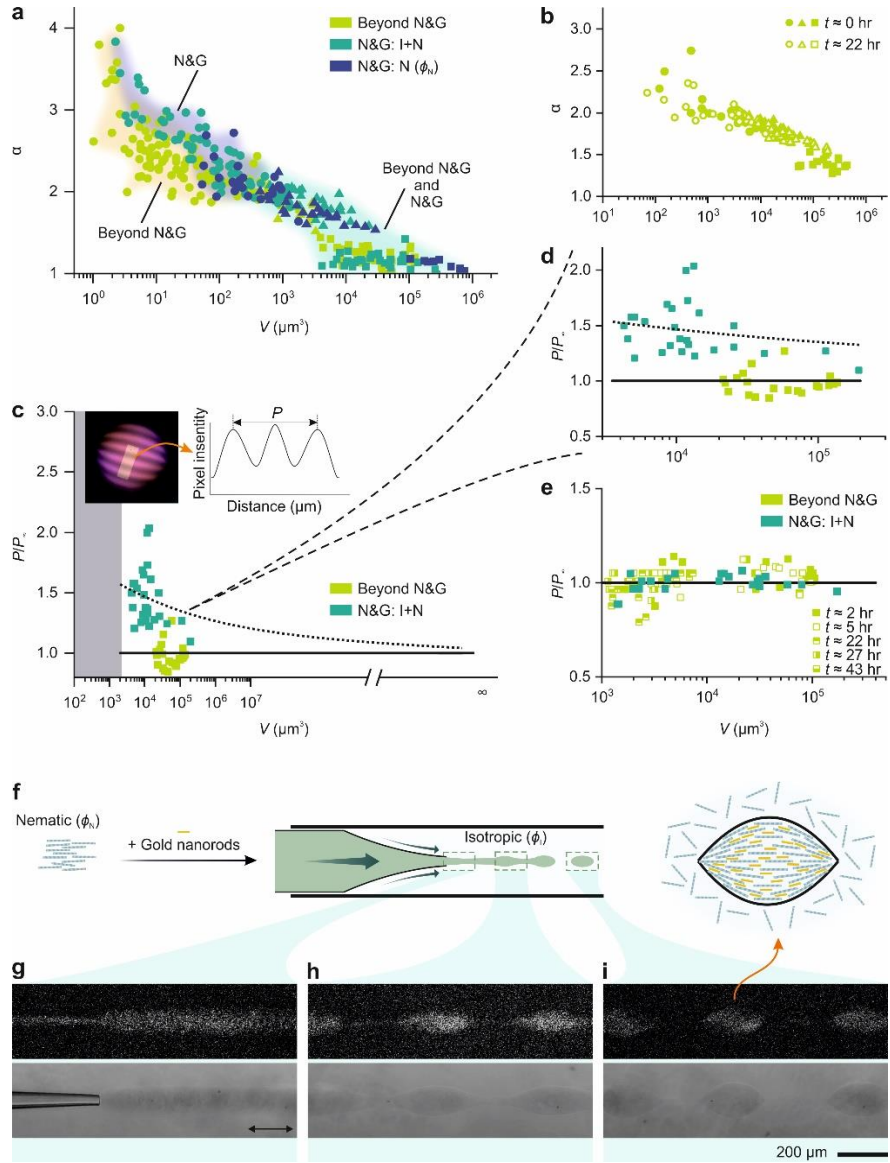

**Fig. 4 | Going beyond N&G reveals unprecedented physical phenomena.** The circle, triangle, and square symbols in the plots represent data points from homogenous, bipolar, and cholesteric tactoids, respectively. Filled symbols denote the tactoids captured right after the relaxation, while unfilled symbols represent long-time measurements. **a**, Nematic-cholesteric phase diagram of liquid crystalline tactoids obtained beyond N&G versus two control experiments following phase separation via N&G (see main text). **b**, Aspect ratio versus volume of homogenous, bipolar and cholesteric BLG tactoids formed beyond N&G and tracked over time. **c**, Pitch $P$ of cholesteric tactoids of BLG obtained via and beyond N&G. While the pitch value decreases with an increase in volume for tactoids formed via N&G, it remains constant for tactoids formed beyond N&G. $P$ is renormalized by the natural pitch of the system $P_\infty$ which is 16.7 µm for N&G measured at full phase separation from a composition falling within I+N region and 46.1 µm for tactoids formed beyond N&G, calculated as an average value of the pitch of the tactoids. **d**, The inset shows the detailed evolution of the pitch with an in increase in tactoids volume. **e**, $P$ of cholesteric droplets from SCNC compared to those obtained the beyond N&G and N&G pathways. Tracking $P$ of the tactoids over long time shows no significant changes. $P_\infty$ is measured to be 15.9 µm for N&G and 14.6 µm for tactoids obtained beyond N&G. **f-i**, Formation of hybrid amyloid fibrils-plasmonic gold nanorods tactoids beyond N&G. Extruding the nematic phase mixed with gold nanorods into the isotropic phase (**f**) results in formation of the hybrid jet (**g**) driven by instability (**h**) breaking to chain of hybrid tactoids (**i**). The images in (**g-i**) are obtained by exciting the samples with linearly polarized light and collecting the confocal fluorescence signal (top row) or the transmitted light (bottom row). The polarization direction of the excitation laser (633 nm) is denoted with black double arrow.

# Supporting Information for

## Disentangling kinetics from thermodynamics in heterogeneous colloidal systems

Hamed Almohammadi[1], Sandra Martinek[1], Ye Yuan[1], Peter Fischer[1], Raffaele Mezzenga[1,2*]

[1]Department of Health Sciences and Technology, ETH Zurich, Zurich, Switzerland
[2]Department of Materials, ETH Zurich, Zurich, Switzerland
*Correspondence to: raffaele.mezzenga@hest.ethz.ch

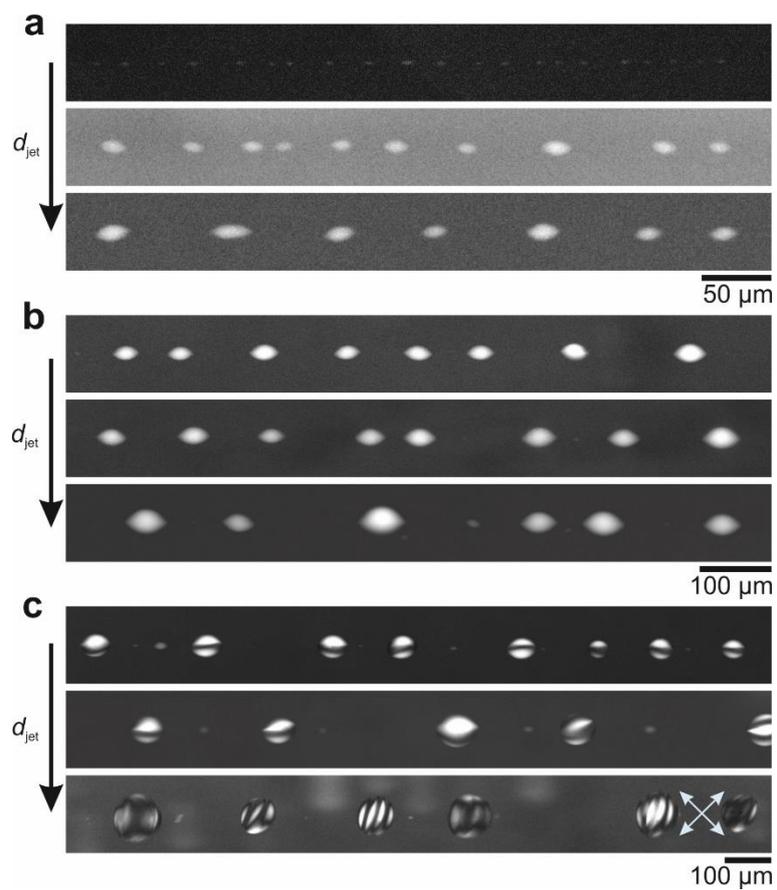

**Supporting Fig. 1 | Formation of nematic and cholesteric BLG tactoids with various size, shape, and internal structures.** Homogenous (**a**), bipolar (**b**) and cholesteric (**c**) tactoids with different volumes are formed by extruding nematic jets with various diameter which achieved by adjusting $q_{in}$ and $q_{out}$. All tactoids are oriented with their long axes in the flow direction, except cholesteric tactoids with four or higher number of bands that are oriented almost perpendicular to flow direction. The crossed arrows represent the crossed polarizers.

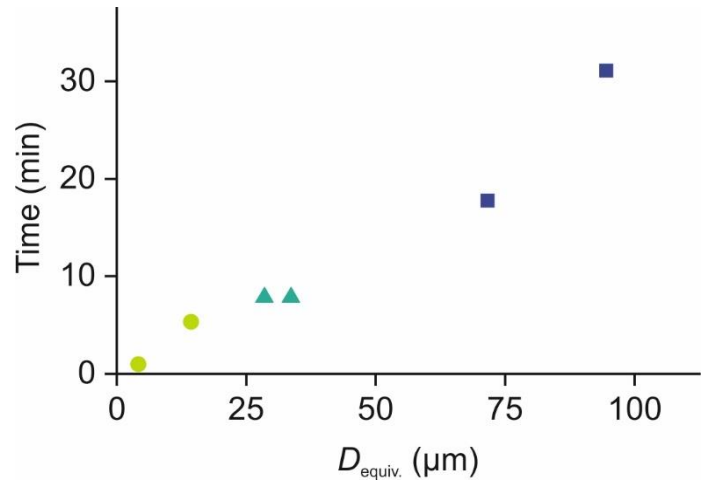

**Supporting Fig. 2 | *Induction time* of the liquid crystalline tactoids formed beyond N&G pathway.** The filled circle, triangle, and square symbols denote homogenous, bipolar, and cholesteric tactoids, respectively. The results show that the *induction time* increases with an increase in the volume of the tactoids and, importantly, it is in the order of the minutes. *Induction time* is defined as the time elapsed from the moment the jet is extruded till when the tactoids reach their equilibrium states (see Fig. 2a).

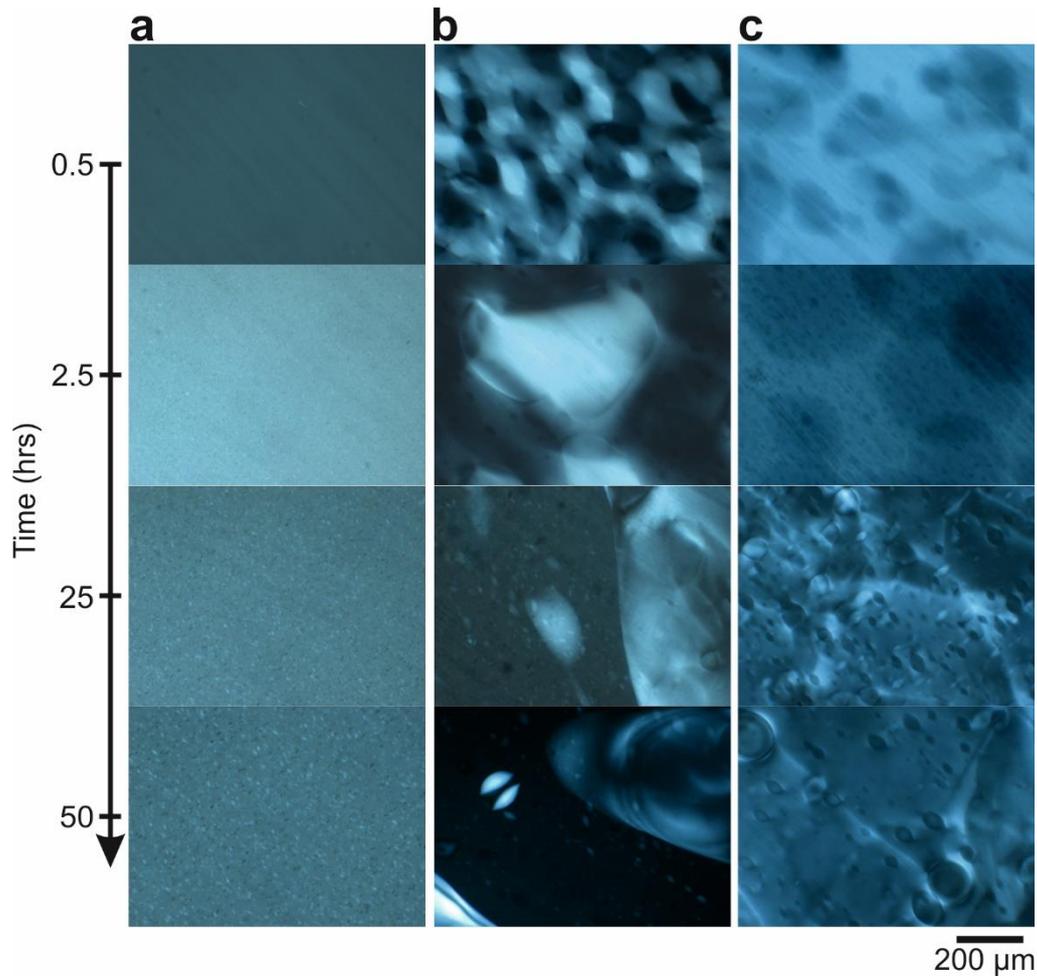

**Supporting Fig. 3 | Classical N&G time for the formation of tactoids.** Three different sets of suspension of BLG with concentrations, within Onsager branches ($\phi_I$ and $\phi_N$), as: (**a**) just above $\phi_I$ at $1.008\phi_I$, (**b**) midpoint of $\phi_I$ and $\phi_N$ at $(\phi_I + \phi_N)/2$, and (**c**) just below $\phi_N$ at $0.996\phi_N$. Within the time frame of observation, cholesteric tactoids are only observed in (**b**), but after two days. In the case of (**a**) mainly homogenous and a few bipolar tactoids are formed within the first two days. At the concentration corresponding to (**c**), mainly negative tactoids are formed with few homogenous and bipolar tactoids that started to appear within 30 mins.

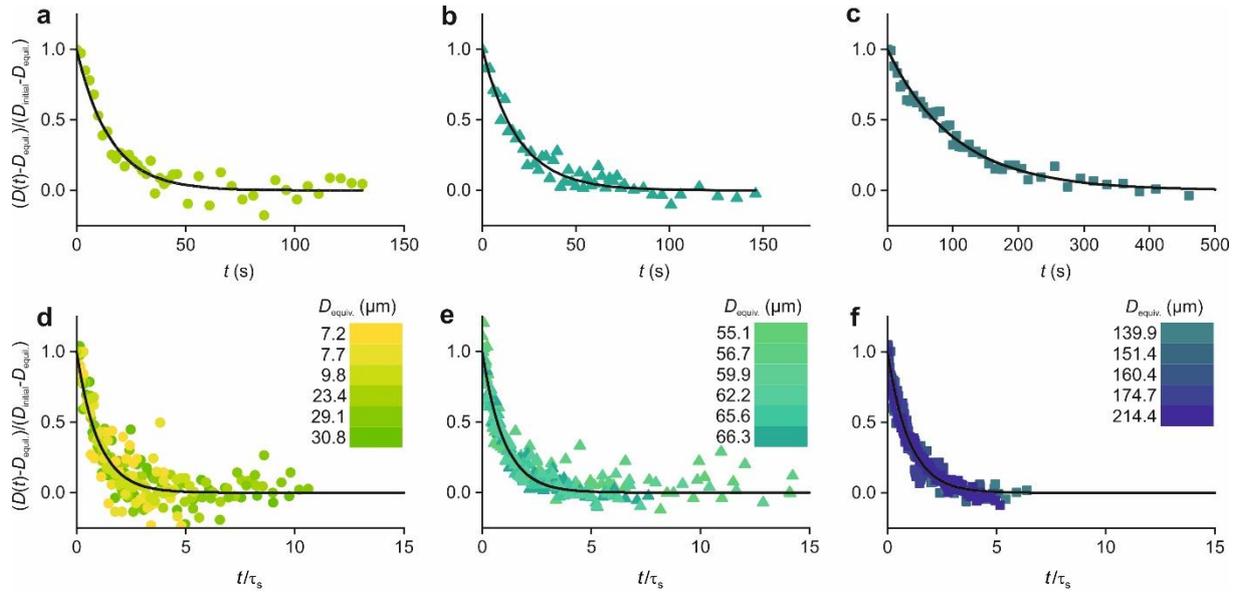

**Supporting Fig. 4 | Shape relaxation behaviour of the tactoids upon breakup of the nematic jet.** The filled circle, triangle, and square symbols illustrate homogenous, bipolar, and cholesteric tactoids, respectively. **a-c**, $(D(t)-D_{equil.})/(D_{initial}-D_{equil.}) = \exp(-t/\tau_s)$, black lines, are fitted to obtain the characteristic shape relaxation time $\tau_s$ (see Ref. 43 for details on $\tau_s$). **d-f**, The results of tactoids of various internal structures, independent from the tactoids volume, follow a universal curve of $(D(t)-D_{equil.})/(D_{initial}-D_{equil.}) = \exp(-t/\tau_s)$, see Fig. 2c.

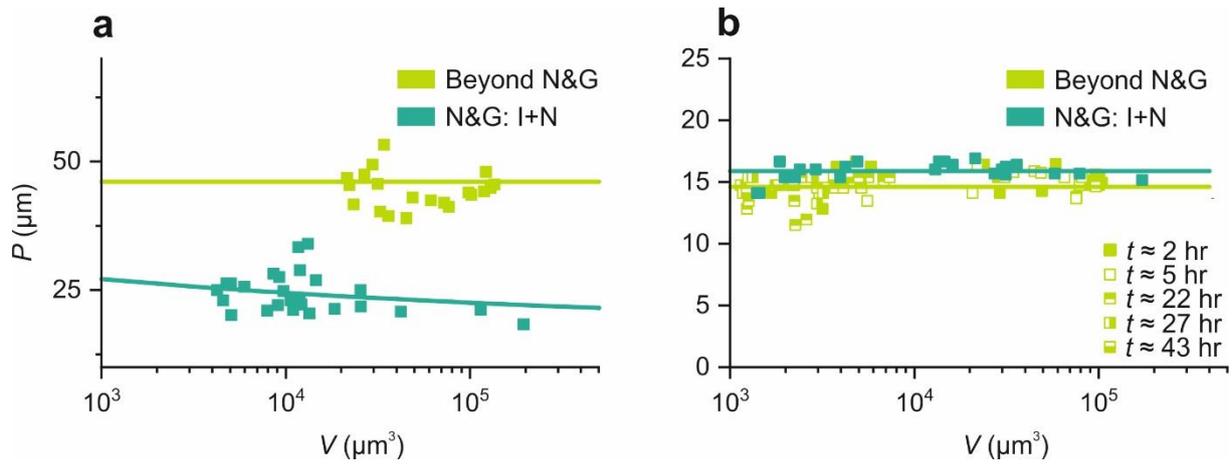

**Supporting Fig. 5 | Comparison of the pitch value of the cholesteric tactoids formed through N&G process and beyond N&G.** The pitch values are from the tactoids of BLG (**a**) and SCNC (**b**). **a**, While the pitch value decreases with an increase in volume for tactoids formed through phase separation via N&G, it stays constant for tactoids formed beyond N&G at 46.1 µm. **b**, $P$ stays constant for tactoids with various volumes formed beyond N&G and via N&G pathways at 14.6 µm and 15.9 µm, respectively. Tracking the $P$ of the tactoids over long time shows no significant changes.

**Supporting videos**

**Supporting Video 1 | Dynamics of the homogenous tactoids with $D_{equiv.}$ = 4.2 ± 0.7 µm formation beyond N&G.** The jet diameter is 2.3 µm. The video was captured under crossed polarizers and the duration of the video is 2.25 min in real time. The white crossed arrows denote the crossed polarizers. The scale bar is 100 µm.

**Supporting Video 2 | Dynamics of the homogenous tactoids with $D_{equiv.}$ = 14.4 ± 1.6 µm formation beyond N&G.** The jet diameter is 7.0 µm. The video was captured under crossed polarizers and the duration of the video is 5.33 min in real time. The white crossed arrows denote the crossed polarizers. The scale bar is 200 µm.

**Supporting Video 3 | Dynamics of the bipolar tactoids with $D_{equiv.}$ = 28.5 ± 3.6 µm formation beyond N&G.** The jet diameter is 12.3 µm. The video was captured under crossed polarizers and the duration of the video is 7.85 min in real time. The white crossed arrows denote the crossed polarizers. The scale bar is 200 µm.

**Supporting Video 4 | Dynamics of the bipolar tactoids with $D_{equiv.}$ = 33.7 ± 4.9 µm formation beyond N&G.** The jet diameter is 18.9 µm. The video was captured under crossed polarizers and the duration of the video is 7.85 min in real time. The white crossed arrows denote the crossed polarizers. The scale bar is 200 µm.

**Supporting Video 5 | Dynamics of the cholesteric tactoids with $D_{equiv.}$ = 71.6 ± 7.8 µm formation beyond N&G.** The jet diameter is 31.8 µm. The video was captured under crossed polarizers and the duration of the video is 17.78 min in real time. The white crossed arrows denote the crossed polarizers. The scale bar is 200 µm.

**Supporting Video 6 | Dynamics of the cholesteric tactoids with $D_{equiv.}$ = 94.5 ± 7.3 µm formation beyond N&G.** The jet diameter is 48.3 µm. The video was captured under crossed polarizers and the duration of the video is 31.66 min in real time. The white crossed arrows denote the crossed polarizers. The scale bar is 200 µm.

**Supporting Video 7 | Dynamics of the negative tactoids formation beyond N&G.** The jet diameter is 55.2 µm. The video is sequence of 3D reconstructed images captured with confocal microscopy. The duration of the video (8.61 min in real time) and the scale bar are provided in the video.